\documentstyle[preprint,aps,accept]{revtex}
\tightenlines
\begin{document}
\draft


\title{Interactions Between Charged Spheres in Divalent Counterion Solution}
\author{Niels Gr{\o}nbech-Jensen and Keith M.\ Beardmore}
\address{Theoretical Division,
               Los Alamos National Laboratory,
               Los Alamos, New Mexico 87545}
\author{Philip Pincus}
\address{Department of Physics,
               University of California,
               Santa Barbara, California 93106}
\date{\today}
\maketitle
\begin{abstract}
We simulate model systems of charged spherical particles in their counterion
solution and measure the thermodynamic pressure and the
pair distribution function from which we derive
effective potentials of mean force. For a system with only electrostatic and
hard core interactions, we investigate the effective potential between two
like-charged spheres in divalent counterion solution as a function of
concentration. We find a strong attractive interaction for high concentration
and a global repulsive effective interaction for dilute systems. The results
indicate a first order phase transition in sphere-counterion density
as a function of global concentration and the effective sphere-sphere
potentials in the dilute (solvated) regime suggest significant density
fluctuations due to short range local minima in the effective energy surface.
Our results arise from a minimal approach model of several recent
experiments on polystyrene latex particles in monovalent counterion solution.
\end{abstract}
\newpage

The effective interactions between charged macromolecules in countercharged
aqueous media have been assumed to be governed mainly by simple screened
Coulomb behavior leading to overall repulsion between like charged objects
\cite{Israelachvili_92,Lindsay_82,Derjaguin_41,Verwey_48}. This notion arises
primarily from mean-field analyses, such as Poisson-Boltzmann equations,
of the statistical counterion distribution, and leads to the well known
Derjaguin-Landau-Verwey-Overbeek (DLVO) purely repulsive effective potential
between charged colloidal spherical macroions \cite{Derjaguin_41,Verwey_48}.
However, recent experiments have indicated that the effective interaction
can indeed be attractive in dense suspensions of, e.g., polystyrene latex
particles in aqueous solutions with monovalent counterions
\cite{Tata_92,Crocker_96,Larsen_97,Sharma_97,Daly_81,Arora_85,Ise_83,Ise_80,Kesavamoorthy_89,Ise_90,Yoshino_88}. These experiments, performed with
very low salt concentrations, have observed inhomogeneous colloid densities
under certain conditions and almost perfectly homogeneous densities under
others. Some of the experiments observe the attractive potentials only when
the colloidal spheres are suspended between glass walls
\cite{Crocker_96,Larsen_97} and it is consequently a speculation whether
the effective attraction is mediated by these walls or if the walls
merely provide a confinement of phase space such that the attraction is
enhanced. Recent advances for investigating the effective interactions have
been made \cite{Stevens_96} by performing direct numerical
simulations of charged colloidal spheres and their monovalent counterions
in confined space and only repulsive interactions were found.

As the present state of theoretical predictions is almost entirely limited
to the trivial repulsive behavior between fractionally screened like-charged
objects, we have carried out simulations of a simple system which exhibits
strong effective short as well as long range attractions between like-charged
spherical objects in their neutralizing counterion solution. This is intended
as helping to construct a successful theoretical approach describing the
complex phase behavior of charged colloidal suspensions.
The purpose of this paper is therefore not to give a direct interpretation of
the recent experiments on polystyrene latex particles, but instead point to
a simple system exhibiting a phenomenon that a theoretical approach should
reproduce.

We present simulations of charged colloidal
spheres and their counterions confined in a cubic box volume $V\equiv L^3$,
where $L$ is the box length in each direction, with
periodic boundary conditions applied such that a finite bulk concentration is
simulated. Finite concentration is important since Coulomb interactions
between spherical
objects ($\sim1/r$) are inferior to entropic repulsion ($\sim\ln{r}$)
at large distances. The equilibrium
distance between any charged spherical objects at finite temperature is
consequently divergent in the limit of $N/V\rightarrow 0$, where $N$ is the
number of spherical objects; thus, the effective interaction is
always repulsive in the dilute limit.
We will focus on a system with divalent counterions in order to clearly
demonstrate the
importance of concentration and correlation effects, which are absent in mean
field treatments leading to, e.g., the DLVO potential.
We demonstrate that the effective potential
between two spheres, as derived from the pair distribution function,
can become attractive in finite concentration.
Our simulations are performed as a minimal approach in that
only long-range Coulomb interactions and short-range volume exclusions
contribute to the evolution of the system.

The model under consideration is given by the Overdamped Langevin equation of
motion for the $i$th particle:
\begin{eqnarray}
\nu_i\dot{\bar{r}}_i & = & -\nabla_i E + \bar{n}_i(t) \; ,
\label{eq:Eq_1}
\end{eqnarray}
where $\bar{r}_i$ is the normalized coordinate of the $i$th particle moving
with a normalized friction coefficient $\nu_i$ (note that neither the
friction coefficient nor the time normalization has any importance for
the equilibrium properties of the system).
The total normalized energy $E$ of the system is given by,
\begin{eqnarray}
E & = & \sum_{i=1}^{N}\sum_{j>i}^N\left(\frac{q_iq_j}{\varepsilon}V(r_{ij})+
W_{hc}(r)\right) \; ,
\label{eq:Eq_2}
\end{eqnarray}
where
\begin{eqnarray}
&& W_{hc}(r) \; = \; \nonumber \\
&& \left\{\begin{array}{ccc}
\frac{A_{ij}}{(r_{ij}-r_{ij}^0)^{12}}- \frac{B_{ij}}{(r_{ij}-r_{ij}^0)^{6}} &
, & (r_{ij}-r_{ij}^0)^{6} < 2A_{ij}/B_{ij} \\
0  & , & (r_{ij}-r_{ij}^0)^{6} \ge 2A_{ij}/B_{ij}
\end{array}\right. \; .
\end{eqnarray}
$A_{ij}$ and $B_{ij}$ are the Lennard-Jones volume exclusion parameters.
Notice that we only consider the short range {\it repulsive} part of the
Lennard-Jones potential.
$N$ is the total number of coordinates (spheres and counterions) and $eq_i$ is
the charge of the $i$th particle.
The normalized Coulomb interaction between charges $eq_i$ and $eq_j$ at a
normalized distance $r_{ij}$ is represented by
$\frac{q_iq_j}{\varepsilon}V(r_{ij})$, with dielectric
constant $\varepsilon$ (for details on exact summation of Coulomb
interactions in orthorhombic systems, see Ref.\ \cite{Jensen_97}).
The thermal equilibrium with the
surroundings, i.e., with the water which is not considered except as a
homogeneous dielectric, is maintained by the noise term $\bar{n}_i(t)$ which is
linked to the dissipation-fluctuation theorem \cite{Parisi_88} by the
uncorrelated white Gaussian distribution,
\begin{eqnarray}
\langle\bar{n}_i(t)\rangle & = & \bar{0} \; \; , \; \;
\langle\bar{n}_i(t)\cdot\bar{n}_j(t')\rangle \; = \;
6\nu_i\frac{k_BT}{E_0}\delta_{ij}
\delta(t-t') \; ,
\label{eq:Eq_3}
\end{eqnarray}
where $\delta_{ij}$ is the Kronecker delta function and $\delta(t-t')$ the
Dirac delta function. Boltzmann's constant is denoted $k_B$, the temperature
is $T$, and the energy is normalized to $E_0$=1 kcal/mol.
The factor, $6$, is due to $\bar{n}$ being a three dimensional vector.
Throughout this paper we have chosen divalent counterions, sphere charges
of $q_{sp}=-10$, $A_{ij}\approx21\cdot$10$^{3}$, $B_{ij}\approx30$, and
$\varepsilon=80$ in order to mimic bulk screening of water.
The hard core parameter,
$r_{ij}^{0}$, is chosen as $r_{ij}^{0}=0$ for counterion counterion
interactions, $r_{ij}^{0}=5.5$ for counterion sphere interactions, and
$r_{ij}^{0}=11$ for sphere sphere interactions.
Each simulation is initiated with random, non-overlapping
positions of the counterions and spheres and a large number of time steps,
of order 10$^8$, are discarded before accumulating the pair distribution
density function, $\rho(r)$, between spheres. In equilibrium, this density
is directly related to the Boltzmann factor,
\begin{eqnarray}
\rho(r) & \sim & \exp\left(-W(r)/k_BT\right) \; ,
\label{eq:Eq_4}
\end{eqnarray}
where $W(r)$ is the effective potential of mean force (PMF).
Inverting this relationship and
normalizing the potential to the value at maximum distance $r=L/2$, we obtain
the derived effective PMF from,
\begin{eqnarray}
W(r) & = & -k_BT\ln\left(\frac{\rho(r)}{\rho(L/2)}\right) \; .
\label{eq:Eq_5}
\end{eqnarray}

Figure 1 displays the PMF between two spheres and their accompanying divalent
counterions in a simulation box of volume $V=L^3$.
The friction coefficients are set to $\nu_{ci}=1$ for counterions and
$\nu_{sp}=15$ for spheres with the accompanying reduced time step $dt=0.005$.
The resulting sphere-sphere PMF, derived over $\sim$10$^8$ time steps,
reveals strongly attractive behavior for small $L$ (lower curves),
but as $L$ is increased, the available volume increases, and the minimum
characteristic for attraction lifts above zero (normalized to the potential at
distance $L/2$) and eventually disappears. The minimum of the PMF, $W_{min}$,
is shown in Fig.\ 2 as a function of the simulated volume. Here we observe
how the minimum (relative to the value at $L/2$) of the effective potential
depends on the volume. At a volume
$\approx (120{\rm \AA})^3 \approx 2\cdot10^3{\rm nm}^3$ the potential at
$L/2$ becomes the global minimum, which clearly indicates a
transition from close packed spheres being preferred in small
volumes to separated spheres being statistically preferred at larger volumes.

However, as is obvious from Fig.\ 1, the spheres can be attracted to each other
at short distances even if the global PMF minimum is at $L/2$. This feature
arises from the much stronger interaction that exists between a counterion
and two closely spaced spheres when compared to the interaction between
a counterion and a single sphere. This then leads to a compact state of
the two spheres with their counterions spatially correlated around them.
The barrier
height (PMF) in energy going from the compact state to the maximum separation
($L/2$) is shown in Fig.\ 3, where we find a significant barrier, i.e., a
local minimum for the compact state, well into the volume regime where the
global minimum of the PMF is at maximum separation.

The PMF potentials shown in figure 1 are in strong contrast to the DLVO
potential, which is given by \cite{Derjaguin_41,Verwey_48}
\begin{eqnarray}
W_{{\rm{DLVO}}}(r) & = &
\frac{k_BT}{E_0}q_{sp}^2\left(\frac{e^{\kappa a}}{1+\kappa a}\right)^2
e^{-\kappa r}\frac{\lambda_B}{r} + W_{hc}(r)\; ,
\label{eq:Eq_DLVO}
\end{eqnarray}
where $a$ is the radius of the spheres and the hard core interaction between two
spheres is given by $W_{hc}$. The two characteristic length scales, $\lambda_B$
and $\kappa$, are given by,
\begin{eqnarray}
\lambda_B & = & \frac{e^2}{4\pi\epsilon_0\epsilon k_BT} \; , \label{eq:Bjerrum} \\
\kappa^2 & = & 4\pi\lambda_B\frac{M}{V}q_{ci}^2 \; , \label{eq:kappa}
\end{eqnarray}
where $M$ is the number of counterions and $q_{ci}$ is the counterion valency.
The DLVO potential is obviously repulsive and monotonically decaying with
distance.

Studying the PMF gives information about the spatial correlations within
the sample simulation, but it does not reveal the overall nature of global
attraction versus repulsion of an ensemble.
In order to study if the system tends to
collapse or expand, we have calculated the thermodynamic pressure,
\begin{eqnarray}
p_T & = & -\left(\frac{\partial {\cal{F}}}{\partial V}\right)_{T,N} \nonumber \\
& = & N\frac{k_BT}{V} 
+ \frac{1}{3V}\langle\sum_{i=1}^N\sum_{j>i}^N\bar{r}_{ij}\cdot\bar{f}_{ij}
\rangle + \frac{\langle U_{el}\rangle}{3V} \; ,
\label{eq:Eq_6}
\end{eqnarray}
where ${\cal{F}}$ is Helmholtz's free energy and $\langle\cdots\rangle$ denotes
a temporal average. The first term on the right hand
side is the {\it thermal} contribution to the pressure,
the second is the contribution from the virial of all {\it hard core}
interactions (\ref{eq:Eq_3}), and the last is the {\it electrostatic}
contribution, $U_{el}$ being the total electrostatic energy of the system.

Figure 4a shows the total pressure, $p_T$, versus the volume for one simulated
sphere and its countreions in a cubic box. In agreement
with the PMFs visualized in figure 1, we find an effective attraction
(negative pressure) for high concentration of spheres and a repulsive
effective interaction (positive pressure) for dilute systems. This behavior
of pressure as a function of concentration indicates a first order phase
transition from a solvated state of spheres at low concentration to a
collapsed state of spheres at high concentration. However, the global pressure
cannot reveal the short range local minima in the PMFs observed in figure 1.
These local minima, which persist far into the dilute limit, will result in
local density fluctuations of the solvated spheres. Note that the negative
pressure is solely due to the electrostatic interactions since both the
kinetic and the hard core contributions to the total pressure are always
positive (see figure 4b, where the three individual pressure contributions are
shown). The total pressure will, in the dilute limit, be entirely dictated
by the thermal contribution (ideal gas) since the finite temperature will
separate all spherical particles and thereby make the hard core and the
electrostatic contributions negligible.

Also the pressure is in contrast to the DLVO analysis. Since the DLVO potential
is purely repulsive, one can only expect positive pressures from such a
treatment. The reason for the possibility of negative pressures in the ``all
particle'' model lies in the overall charge neutrality of the system. It is
clear that a charge neutral system composed of oppositely charged components
is self-contracting at low temperatures (e.g., the sodium-chloride crystal) due
to strong spatial correlations between the charged objects. However, the DLVO
analysis only accounts for the counterion charge if it contributes to the
screening of a sphere, and thus, the DLVO treatment does not maintain the
overall charge neutrality of the system.

We have simulated a system of charged spheres in divalent counterion solution
and demonstrated a very strong attractive effective potential between the
like-charged spheres. The origin of the attraction lies in the strong
correlation effects between counterions when the spheres are in a compact
conformation. This simple result is in contrast to the effective (and
always repulsive) DLVO potential, which is developed without considering the
counterions as particles. However, our results are in agreement with other
recent simulations of like-charged rods in divalent counterion solution
\cite{Jensen_97_2} and theoretical considerations regarding like-charged plates
\cite{Rouzina_96}. Our simple simulations demonstrate that a successful
theory explaining the complicated phase diagram of complex charged fluids
needs to account for the correlated behavior of counterions, the global
charge neutrality leading to the tensile pressure in figure 4, and the
relationship between the sphere radius and the simulated volume responsible for
the statistical distribution of counterions.

It is important to emphasize that
our simulations are not meant to characterize real experiments quantitatively.
The minimal approach model presented in this paper cannot account for any
atomic or molecular detail that may be important for experiments. For example,
including the effects of the solvent (water) as friction, noise, and bulk
screening is a dramatic simplification of the true discrete nature of water
molecules, which exhibit their discrete nature within the hydration shell.
Also, neglecting the complexity of the dielectric mixture of water and spheres
prevents our model from correctly simulating the electrostatic problem. Our
simulations therefore address the validity of present theories,
such as DLVO, which attempt to describe the minimal approach model described
in this paper.

This work was performed under the auspices of the U.S.\ Department of Energy,
supported in part by National Science Foundation grant DMR-9708646 and in part
by funds provided by the University of California for the conduct of
discretionary research by Los Alamos National Laboratory.



\vspace*{-0.2 in}

\begin{figure}
 
\vspace{4.0 in}
 
\hspace{6.0 in}
\includegraphics{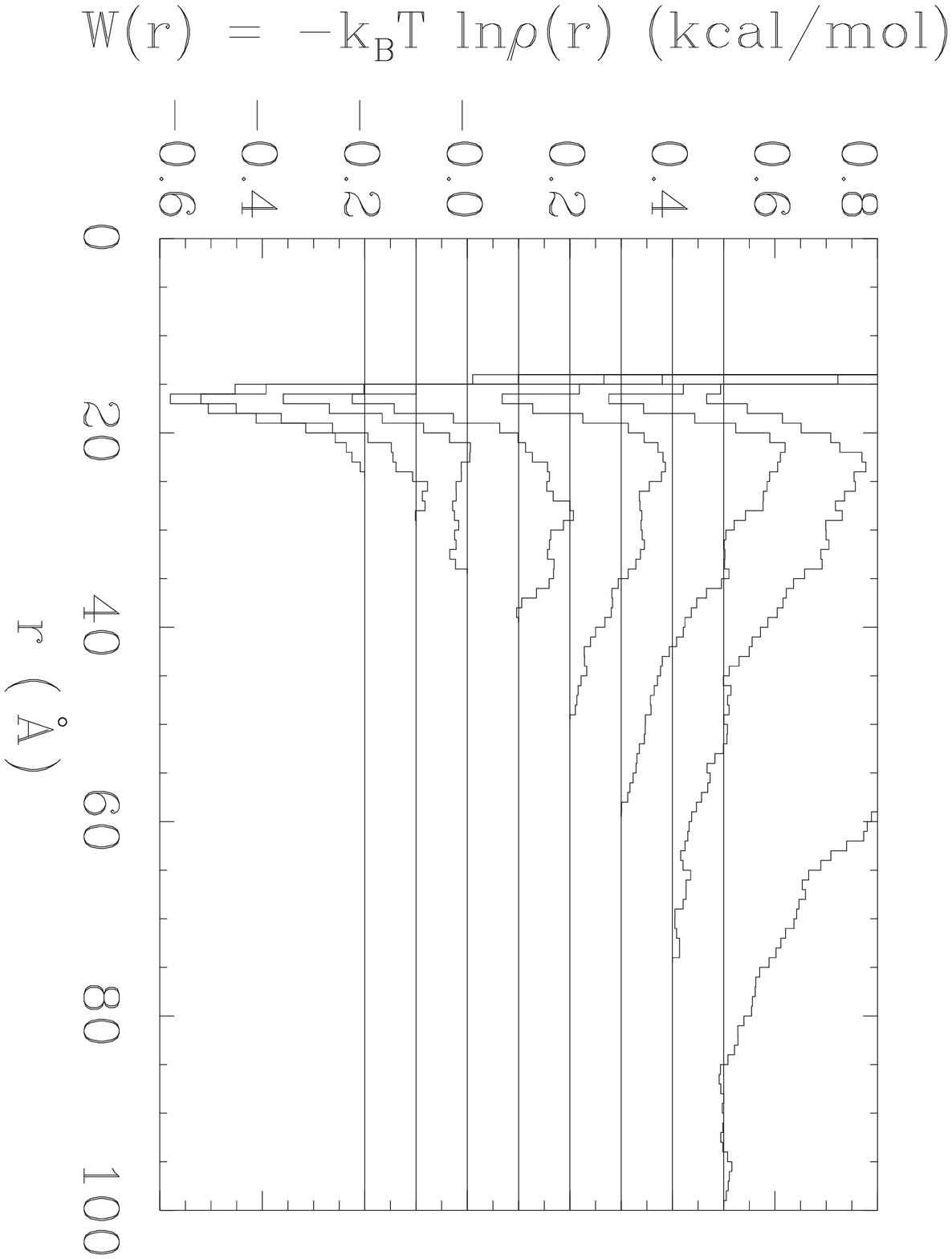}

\caption{Potentials of Mean Force for different simulated volumes.
The potentials are set to zero at maximum separation, $L/2$, and the
different potentials are off-set vertically in order to better distinguish
the cases. Potentials of Mean Force are shown for (from the top):
$L=200${\AA}, $L=150${\AA}, $L=120${\AA}, $L=100${\AA}, $L=80${\AA},
$L=70${\AA}, $L=60${\AA}, and $L=50${\AA}}.
\end{figure}

\vspace*{-0.3 in}

\begin{figure}
 
\vspace{4.0 in}
 
\hspace{6.0 in}
\includegraphics{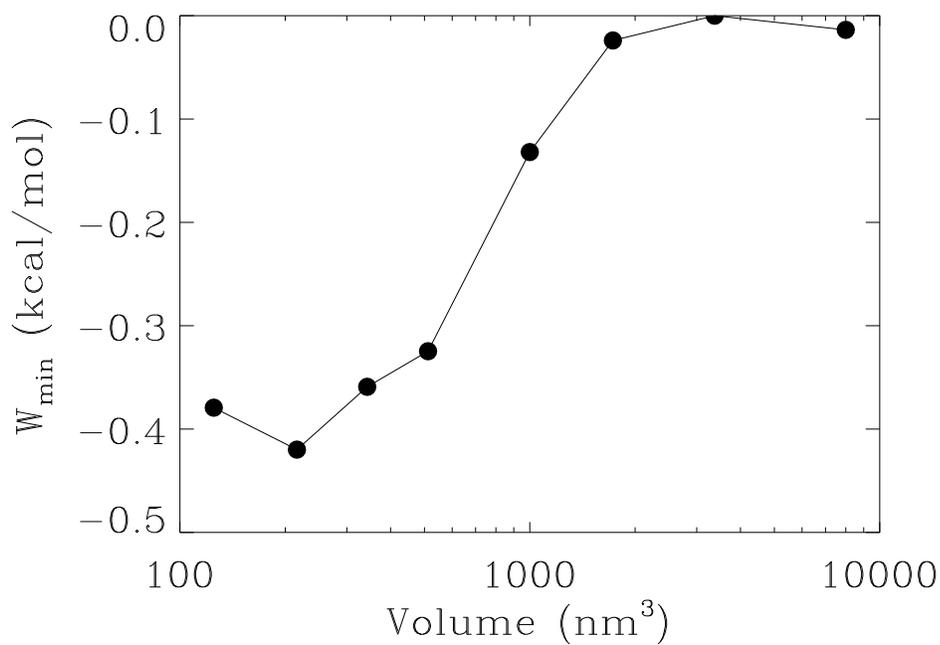}
\caption{Minimum of the potentials of mean force shown in Fig.\ 1 versus
the simulated volume, $V=L^3$.}
\end{figure}

\vspace*{-0.3 in}

\begin{figure}

\vspace{4.0 in}
 
\hspace{6.0 in}
\includegraphics{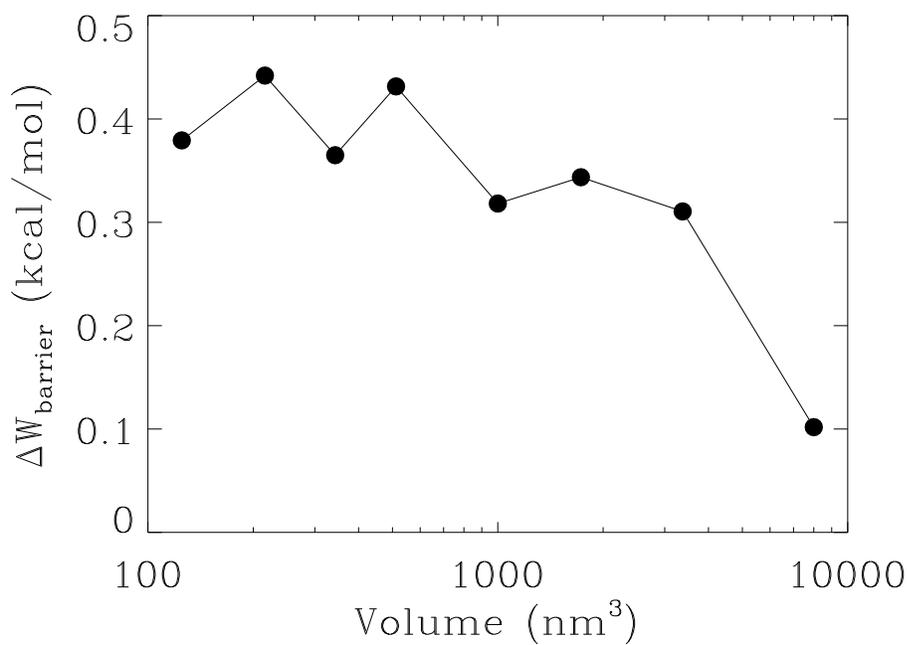}
\caption{Energy barrier height from the compact state ($r\approx 16$ in figure
1) to the local PMF maximum ($r\approx 22$ in figure 1) as a function of the
simulated volume.} 
\end{figure}

\vspace*{-0.3 in}

\begin{figure}

\vspace{4.0 in}
 
\hspace{6.0 in}
\includegraphics{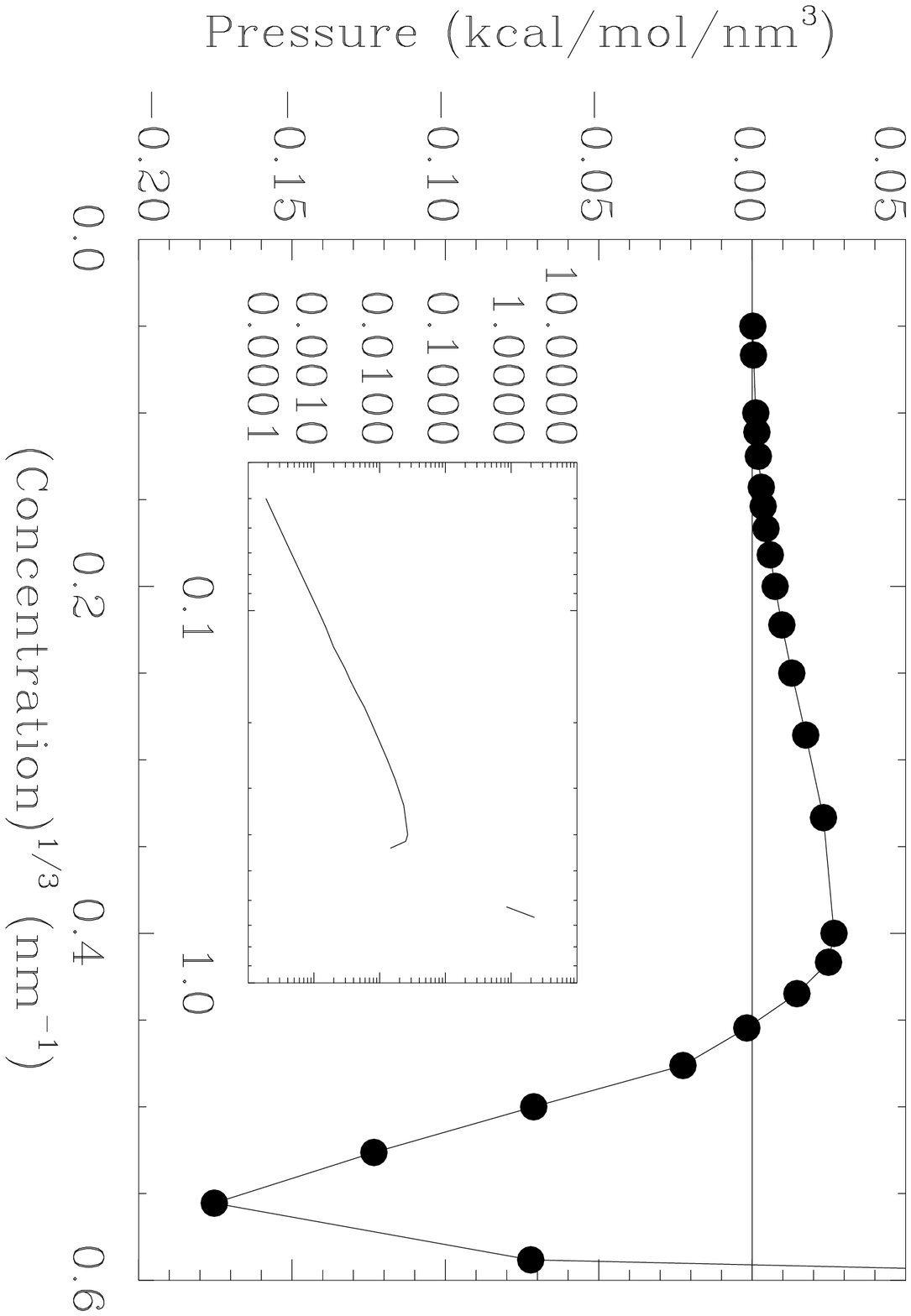}

\vspace{4.0 in}
 
\hspace{6.0 in}
\includegraphics{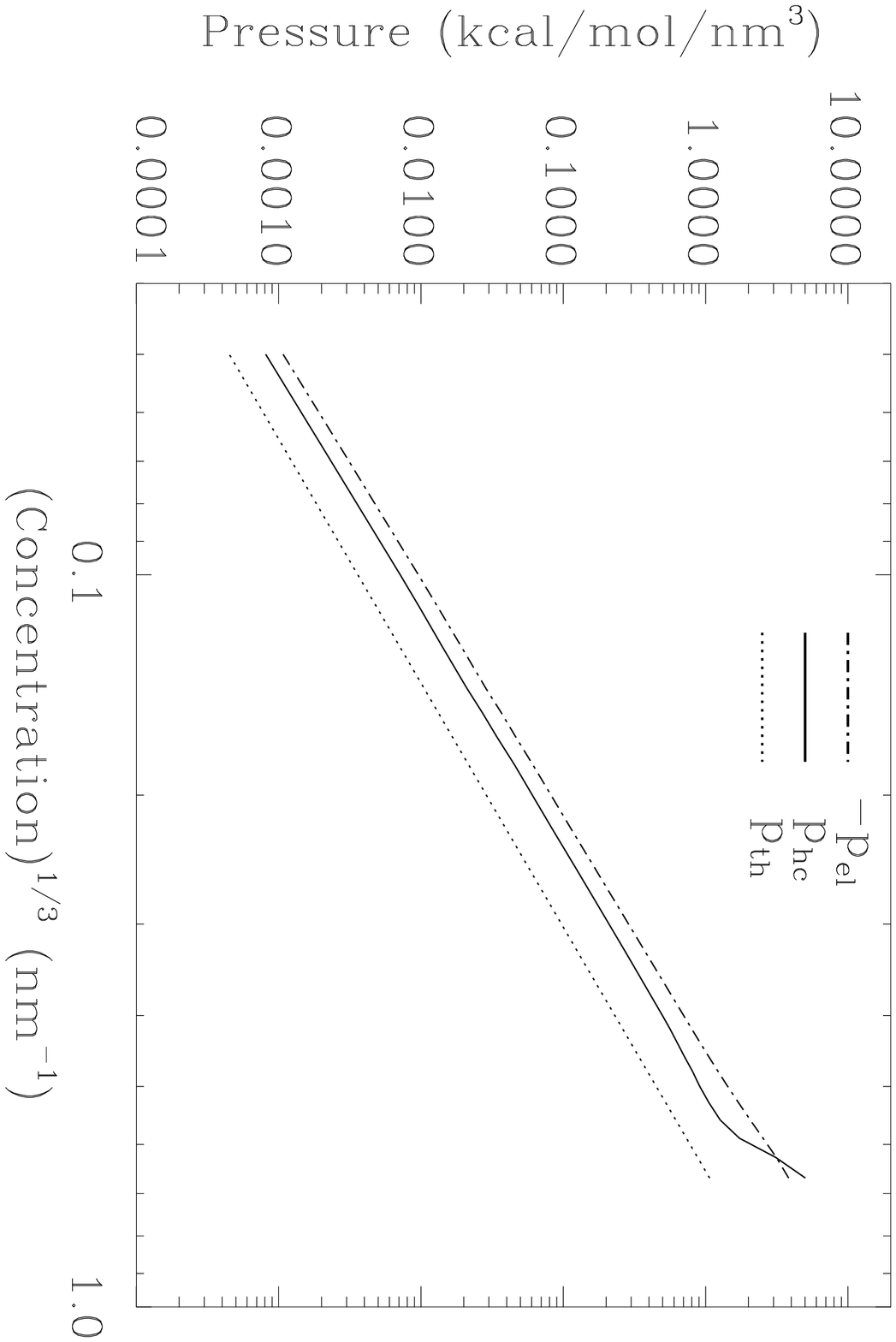}
\caption{(a)
Magnitude of thermodynamic pressure of a system of one sphere and its
counterions in a cubic box versus simulated volume. The inset shows the same
data in a log-log representation.
(b) Magnitude of the three contributions to the pressure shown in (a);
thermal (dotted); hard core (solid); and the negative electrostatic (dash-dot).
}
\end{figure}
\end{document}